\documentclass[aps,prl,twocolumn,footinbib,superscriptaddress,floatfix]{revtex4-2}

\usepackage{graphicx}% Include figure files
\usepackage{dblfloatfix}
\usepackage{amssymb}
\usepackage{amsmath}
\usepackage{times}
\usepackage{bm}
\usepackage{xcolor}

\begin{document}
	
\title{Lyapunov vectors and excited energy levels of the directed polymer in random media}

\author{Enrique Rodr\'iguez}
\affiliation{Departamento de Matem\'aticas, Universidad Carlos III Madrid, Spain}

\affiliation{Instituto de F\'isica de Cantabria (IFCA), CSIC-Universidad de
	Cantabria, 39005 Santander, Spain}

\author{Juan M. L\'opez}
\email{lopez@ifca.unican.es}
\affiliation{Instituto de F\'isica de Cantabria (IFCA), CSIC-Universidad de
	Cantabria, 39005 Santander, Spain}

\date{\today}

\begin{abstract}
The scaling behavior of the excited energy levels of the directed polymer in random media is analyzed numerically. We find that the spatial correlations of polymer energies scale as $\sim k^{-\delta}$ for small enough wavenumbers $k$ with a nontrivial exponent $\delta \approx 1.3$. The equivalence between the stochastic-field equation that describes the partition function of the directed polymer and that governing the time evolution of infinitesimal perturbations in space-time chaos is exploited to connect this exponent $\delta$ with the spatial correlations of Lyapunov vectors reported in the literature. The relevance of our results for other problems involving optimization in random systems is discussed. 
\end{abstract}

%\pacs{Valid PACS appear here}

\maketitle

Nonlinear spatially extended dynamical systems often exhibit space-time chaos~\cite{eckmann,ott,PikoPolibook}. For a system with $N$ degrees of freedom, according to Osedelets ergodic theorem~\cite{oseledec}, there exists a set of $L$ real numbers $\lambda_1 > \lambda_2 > \cdots > \lambda_N$, the characteristic Lyapunov exponents (LEs), that measure the exponential rates of separation or convergence of nearby trajectories and provide an important tool to characterize chaos in nonlinear dynamical systems. LEs are associated with certain special directions in tangent space, the so-called covariant/characteristic Lyapunov vectors (CLVs) $g_1(\mathbf{x},t), g_2(\mathbf{x},t), \cdots, g_N(\mathbf{x},t)$. CLVs have shown to be crucial to fully characterize many aspects of chaotic behavior in extended systems (see~\cite{PikoPolibook} for a recent review), including the role of hydrodynamic modes~\cite{GoodHLM,Romero-Bastida2012}, chaos extensivity~\cite{takeuchi11} and hyperbolicity~\cite{yang09,Kuptsov2012}, time-delayed systems~\cite{pazo_lopez10}, LE fluctuations~\cite{plp13,plp16}, as well as for initializing ensembles in forecasting applications~\cite{pazo10}. This outburst of activity was mainly driven by the discovery of numerical algorithms to effectively calculate the full set of CLVs~\cite{wolfe07,ginelli07,Kuptsov2012a} in spatially extended systems of arbitrary complexity and large size.

It was early on noticed that CLVs have rather peculiar localization properties and exhibit scale-invariant correlations~\cite{szendro07,pazo08}. These spatial correlations (to be defined below) are characterized by a power spectral density $S(k) \sim k^{-\delta}$ with a universal critical exponent $\delta \approx 1.2$, which seems to describe long wavelength correlations for many different spatially extended systems that exhibit dissipative chaos. This new scaling exponent has been shown to be crucial to explain the universal scaling of LE fluctuations in space-time chaos~\cite{plp13}. The origin of the new exponent and its significance in other contexts of physics are still open questions.

In this Letter we show that the critical exponent $\delta \approx 1.2-1.3$ also appears in a very different, seemingly unrelated context: the scaling properties of the excited energy levels of the directed polymer in random media (DPRM). The DPRM problem plays a central role in condensed matter physics~\cite{Halpin-Healy.Zhang_1995} with mappings or connections that include the problem of finding optimal paths in random media~\cite{Kardar.Zhang_1987,Fisher1991}, fracture cracks~\cite{Hansen.etal_1991,Picallo2009}, surface growth~\cite{Kardar1987,Halpin-Healy_1989,Calabrese2011}, random matrix theory~\cite{Borodin2016,Corwin2012,Halpin-Healy2015,Takeuchi2018}, among many others. Since the DPRM is, up to some extent, a problem amenable to analytical treatment~\cite{Kardar1987,dotsenko_2000,Calabrese2011,Borodin2016,Dotsenko2017,Dotsenko_2022}, our results have the potential to lead to a mathematical understanding of the origin of this elusive exponent.

\paragraph{LVs in spatially extended systems.-} Given a dynamical system described by the state variable $u(\mathbf{x}, t)$ at position $\mathbf{x} \in \mathbb{R}^d$ and time $t$, 
Lyapunov analysis~\cite{PikoPolibook} focuses on the evolution of a random infinitesimal perturbation $\Upsilon(\mathbf{x},t=0) := \delta u(\mathbf{x},0)$. Both LEs and CLVs fully characterize tangent space dynamics and the geometry of stretching and contracting volumes in phase space. It is important to note that (almost) any random initial perturbation $\Upsilon(\mathbf{x},t)$ will align with the main CLV,  $\Upsilon(\mathbf{x},t) \propto g_1(\mathbf{x},t)$, exponentially fast, since other stretching directions are exponentially less amplifying in comparison with the first~\cite{wolfe07,ginelli07,szendro07,pazo08}. Only if the initial perturbation was ideally set to lie in the $n$-th CLV subspace it would remain aligned with that direction (numerically this idealization is limited by computer precision, for obvious reasons). In other words, CLVs for $n>1$ are unstable solutions of the tangent space dynamics and are physically non accessible but by specially designed algorithms. We shall recall this observation later when we discuss DPRM excited energy levels.   

CLVs are strongly localized in space~\cite{pik98}, at least for directions associated with positive (expanding) LEs. This suggests to study the logarithm of the CLV amplitude as the relevant quantity. In fact, it is customary to define the $n$-th CLV surface $h_n(\mathbf{x} ,t) := \ln |g_n(\mathbf{x} ,t)|$ and calculate spatial correlations of $h_n$~\cite{pik94,szendro07,pazo08}. Numerical studies in a variety of spatially extended dynamical systems (including several coupled-map lattices, coupled symplectic maps, continuous time systems, etc) have shown that CLV surfaces exhibit seemingly universal correlations for chaotic dissipative systems~\cite{pik94,szendro07,pazo08,pazo_lopez10,plp13}. The power spectral density for the $n$-th CLV surface scales as a power law $S_n(k) = \lim_{t\to\infty} \langle |\hat h_n(k,t)|^2\rangle \sim k^{-\delta}$. For the main LV, $n=1$, one finds invariably $\delta = 2$ in $d=1$~\cite{pik94,pik98}. However, for CLVs with $n>1$ numerical simulations have shown a crossover at long wavelengths to a power law with exponent $\delta \approx 1.2$, with a crossover length scale that decreases as $n$ increases~\cite{szendro07,pazo08}. This asymptotic exponent seems to be robust for different systems~\cite{plp13}.

In the spirit of statistical field theory, the evolution of infinitesimal perturbations in spatially extended chaotic systems can been generically described by the heat equation with multiplicative noise~\cite{pik98,szendro07,pazo08}. Specifically, in the hydrodynamic limit we expect the statistical properties of perturbations to be described by an effective theory:
\begin{equation}
\partial_t\Upsilon(x,t) =   \partial_{xx}\Upsilon + \zeta(x,t) \Upsilon,
\label{mult_noise_eq}
\end{equation}
where the diffusion term describes spatial relaxation of local perturbances, while $\zeta(x,t)$ is a multiplicative Gaussian white-noise term that takes into account, in a coarse-grained fashion, the seemingly random fluctuations in amplification/contraction effects of nonlinearities along the trajectory. Such an effective Landau-type theory is expected to capture the universal statistical properties, in particular the scale-invariant spatio-temporal correlations, of tangent space vectors. Similar arguments, for instance, lead to an effective and very successful theory for the dynamics of the synchronization error in coupled chaotic systems~\cite{Ahlers2002} by including a nonlinear term $-\kappa\Upsilon(1-\Upsilon)$ in Eq.~(\ref{mult_noise_eq}).

The effective theory immediately provides an explanation of the correlation scaling for the 1st CLV. Indeed, from Eq.~(\ref{mult_noise_eq}), the surface $h(x,t) := \ln |\Upsilon(x,t))|$ obeys the ubiquitous Kardar-Parisi-Zhang (KPZ) equation~\cite{Kardar.etal_1986} for interfacial dynamics:
\begin{equation}
\partial_t h(x,t) =  \partial_{xx}h + (\partial_x h)^2 + \zeta(x,t),
\label{KPZ}
\end{equation}
which exhibits scale-invariant solutions with surface correlations $\langle |\widehat h(k,t)|^2\rangle \sim k^{-2}$ in 1D~\cite{Kardar.etal_1986}. Since any random perturbation aligns with the main CLV, $\Upsilon(x,t) \propto g_1(x,t)$, the associated surfaces simply differ by a constant and both satisfy KPZ equation, therefore $\delta = 2$ for the 1st CLV.

Actually, the effective field theory (\ref{mult_noise_eq}) should also describe any perturbation evolving in tangent space, including the CLVs for $n>1$. However, as discussed above, CLVs for $n>1$ are not typical solutions of the tangent space dynamics-- namely, its basis of attraction has zero measure-- and they are {\em saddle-point} solutions of the tangent space equations. To be concrete, a random perturbation that is set to be initially in the subspace spanned by the $n$-CLVs for $n = 2, 3, \cdots$ will align with the 2nd CLV, a random perturbation that is set to be initially in the subspace spanned by the $n$-CLVs for $n = 3, 4 \cdots$ will align with the 3rd CLV, and so forth. This reasoning leads to the conclusion that the scaling properties of the $n$-th CLV could be also extracted from Eq.~(\ref{mult_noise_eq}) after the removal of the dominant components. In the following we will show how this can be expressed in terms of the mapping  of Eq.~(\ref{mult_noise_eq}) to the DPRM problem.

\paragraph{Directed polymer in 1+1 dimensions.-} Consider a directed polymer growing in a disordered environment given by the quenched random potential $\zeta(x,t)$ that is uncorrelated. The starting point of the polymer is fixed at $(0,0)$ and the end point is left free. The position of the polymer head is $(x,t)$, where $t$ is the growth direction. 
In the continuum limit the partition function of all paths that connect $(0,0)$ with $(x,t)$ is given by~\cite{Halpin-Healy.Zhang_1995}
\begin{equation}
Z(x,t) = \int_{(0,0)}^{(x,t)} {\cal D}[x(s)] \; e^{-\int_o^t ds \{(\frac{dx(s)}{ds})^2 + \zeta[x(s),s]\}},
\nonumber
\end{equation}  
which following Feynman-Kac formula, it can be shown that
\begin{equation}
\partial_t Z =  \partial_{xx}Z + \zeta(x,t) Z.
\label{partition}
\end{equation}
The free energy of the directed polymer that ends at $(x,t)$ is ${\cal F}(x,t) \propto -\ln Z$ and according to Eq~(\ref{partition}) obeys KPZ scaling. This is all well known in the context of DPRM theory and, indeed, has provided a basis to obtain the exact solution of KPZ in 1D~\cite{Calabrese2011}.

The final states of the polymer are directed paths that minimize the free energy ${\cal F}(x,t)$. Those minimal energy paths are realizations of the ground state, which exhibits sample to sample fluctuations due to random disorder. Note that there is a one to one correspondence with the problem of the LVs: the free energy of the ground state corresponds to the 1st LV surface with the appropriate change of language, $Z \leftrightarrow \Upsilon$ and ${\cal F} \leftrightarrow -h$.

In the following we will study by means of numerical simulations the statistics and correlations of the excited energy states of the DPRM, {\it i.e.} polymer paths whose energy is above the ground state. 

We focus on the zero temperature limit, as we are interested in the strong-coupling limit~\cite{Kardar.Zhang_1987}. We consider the usual setup where a one-dimensional directed polymer grows on a $45^\circ$ rotated square lattice. The starting point of the polymer is fixed at $(0,0)$ and the end point is left free. Directed paths grow along the bonds of the lattice by growing in one unit at each time step. The position of the polymer head is $(x,t) \in \mathbb{Z}\times\mathbb{N}$, where $t$ is the growth direction. The polymer evolves in a quenched disorder background that is implemented by assigning an uncorrelated randomly distributed number $\zeta(x,t)$ to each site of the lattice. At zero temperature the free energy is just the total energy and the ground state is the path whose energy is the minimum over all paths $\gamma_t$ that grow from $(0,0)$ up to time $t$:
\begin{equation}
E_0(t) = \min_{\gamma_t} \sum_{(x,\tau) \in \gamma_t} \zeta(x,\tau),
\label{groundstate}
\end{equation}
where $E_0(t)$ is the ground state energy with $x \in [-t,t]$ and $\tau \in [0,t]$. At any given time $t$ the polymer energies can be computed by means of a transfer matrix recurrence relation~\cite{Derrida1983,Huse1985}:
\begin{align}
E(x,t) = \min[E(x-1,t-1)+ \zeta(x-1,t-1), \nonumber \\
E(x+1,t-1)+ \zeta(x+1,t-1)], \nonumber
\end{align}
which gives the energies of all minimal paths that connect $(0,0)$ to $(x,t)$. The ground state is then computed by taking the minimum of the energies over all end points, $E_0(t) = \min_x E(x,t)$, that is Eq.~(\ref{groundstate}). From all 
the minimal energy paths of length $t$ only one is the ground state, say the one ending at $x=x_0(t)$, which exhibits sample to sample fluctuations due to disorder. The scaling properties of the ground state are known to be related with KPZ exponents and one has that the sample to sample energy fluctuations scale with polymer length as %$\sigma_{E,0}(t)=Var[E_0(t)] \sim t^{2 \beta}$, 
$\sigma^2_{E,0}(t) = \langle E_0^2\rangle - \langle E_0 \rangle^2 ~\sim t^{2\beta}$
while the position of the end-point is 
%$\sigma_{x,0}(t) = Var[x_0(t)] \sim t^{2/z}$, with $\beta = 1/3$ and $z=3/2$, 
$\sigma^2_{x,0}(t) = \langle x_0^2\rangle - \langle x_0 \rangle^2 ~\sim t^{2/z}$
as corresponds to KPZ universality class~\cite{Halpin-Healy.Zhang_1995}. 

The remaining paths, ending at other positions, correspond to excited energy states, which have received very little attention in the literature mainly due to the fact that they are physically irrelevant since they are not accessible at $T=0$. These are the states we are interested in.

%%%%%%%%%%%%%%%%%%%%%%%%%%%%%%%%%%%%%%%%%%%%% FIG 1
\begin{figure}
	\centerline{\includegraphics[width=\columnwidth]{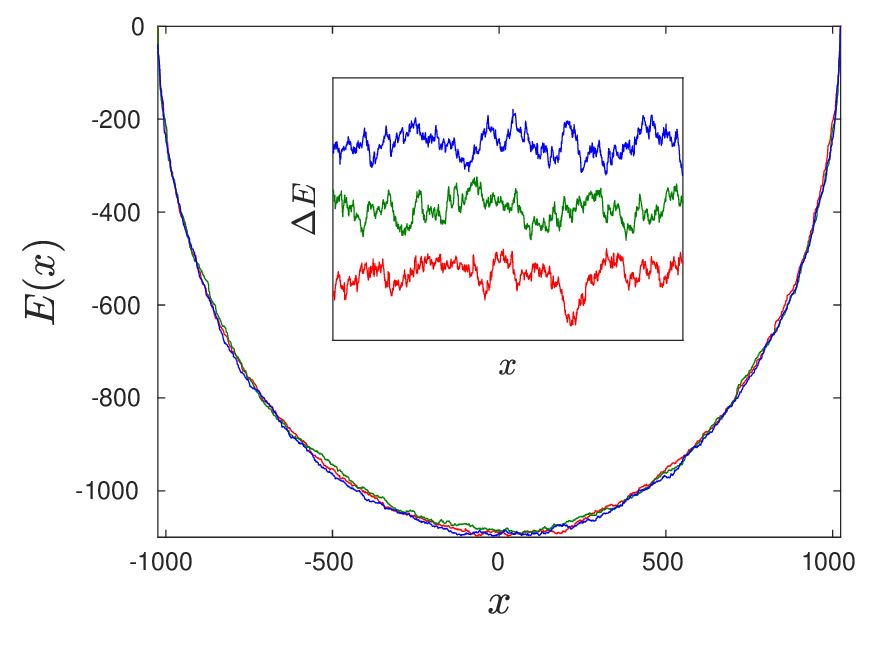}}
\caption{Typical energies $E(x,t)$ for three disorder realizations of a polymer of length $t=2048$ are shown. The ground state fluctuates around $x=0$, while higher energy levels follow a semielliptic profile. The inset shows the energy fluctuations defined as in Eq.\ \eqref{eq:deltae}.}
\label{fig:profile}
\end{figure}
%%%%%%%%%%%%%%%%%%%%%%%%%%%%%%%%%%%%%%%%%%%%%

\begin{figure}
	\centerline{\includegraphics[width=\columnwidth]{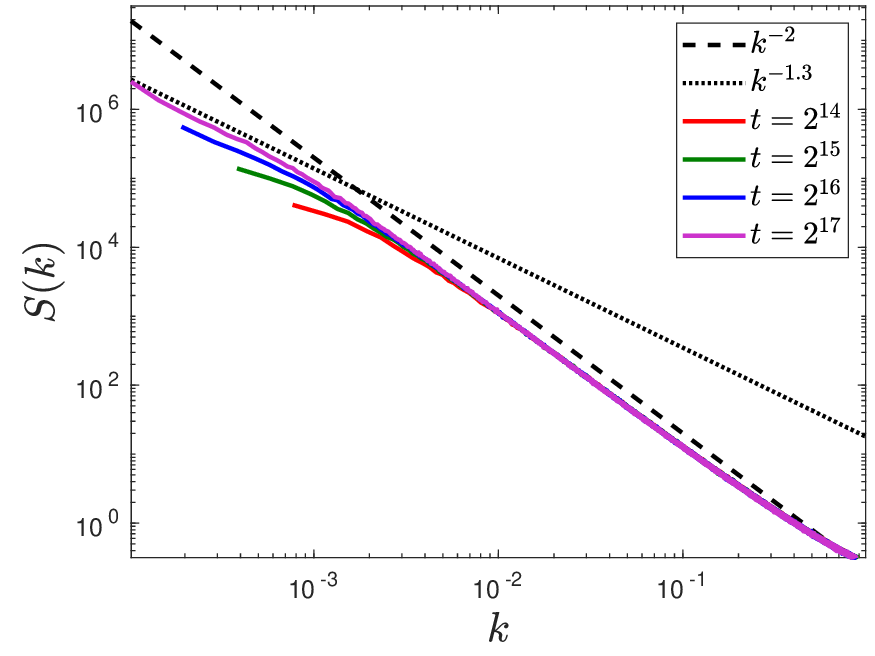}}
	\caption{Power spectral density $S(k,t)$ of energy fluctuations for polymers of lengths $t=2^{14}$ to $2^{17}$. Data are averaged over $10^4$ noise realizations. Dash and dotted lines are guides to the eye corresponding to the exponent $\delta=2$ and $\delta=1.3$, respectively.}
	\label{fig:Sk}
\end{figure}

In Fig.~\ref{fig:profile} we show an ensemble of energy profiles $E(x,t)$ for polymers of length $t=2048$ and different disorder realizations of the DPRM at $T=0$. Similar profiles are obtained for any length $t$. One can see that the energy fluctuates around a semielliptic profile, $ \tilde{E}(x,t) $, that satisfies  $\epsilon(\tilde{E} - a) ^2 + (x-b)^2 = R^2$, with $\epsilon$, $a$, $b$ and $R$ are $t$ dependent parameters. For polymers of length $t$ we define the fluctuation around the energy profile
\begin{equation}\label{eq:deltae}
	\Delta E(x,t) := E(x,t) - \tilde{E}(x,t)
\end{equation}
and calculate the spatial correlations of $\Delta E$ for very long polymers. The parameters that define $\tilde{E}$ are fitted for each disorder realization using the central half domain $[-t/2,t/2]$ in order to avoid the inaccuracies that occur close to the system edges, where the slope of the semielliptic profile diverges. In Fig.~\ref{fig:Sk} we plot the power spectral density $S(k,t) = \langle |\widehat{\Delta E} (k,t)|^2\rangle$, where hat denotes spatial Fourier transform, for polymers of lengths $t=2^{14}, 2^{15}, \cdots, 2^{17}$. One can clearly see a crossover from $k^{-2}$ scaling at short distances to $k^{-1.3}$ as we probe correlations at longer distances. Note that, given the energy profile, spatially close points $x$ correspond (on average) to close energy levels. At short distances ({\it i.e.} small energy gap) the energy fluctuation of excited states appears to be governed by the KPZ universality class, as occurs for the ground state. In contrast, what Fig.~\ref{fig:Sk} shows is that correlations between energy levels with a large gap belong to a different universality class.
%%%%%%%%%%%%%%%%%%%%%%%%%%%%%%%%%%%%%%%%%%%%% FIG 2
%%%%%%%%%%%%%%%%%%%%%%%%%%%%%%%%%%%%%%%%%%%%%

\begin{figure}
	\centerline{\includegraphics[width=\columnwidth]{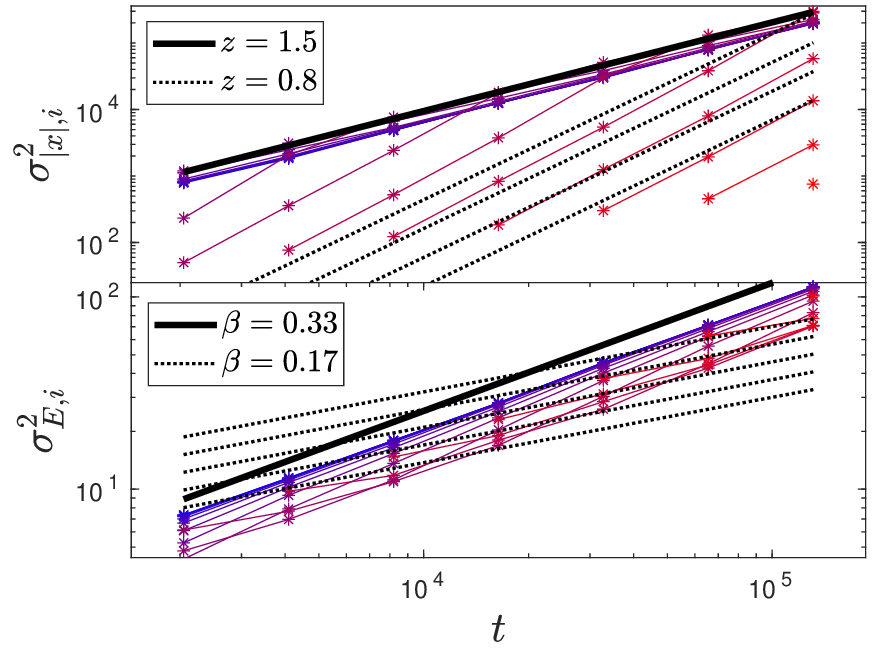}}
	\caption{Time evolution of $\sigma_{E,i}(t)$ and $\sigma_{|x|,i}(t)$ for different energy levels $i=1,2,4,8,...,16384$ (color graded from blue to red) for a polymer of length t=16384. Data are averaged over $10^4$ noise realizations. Straight and dotted lines are guides to the eye for the corresponding values of $z$ and $\beta$ (top and bottom panels) for KPZ scaling and our estimations for high energy levels, respectively.}
	\label{fig:betazeta}
\end{figure}

We have also studied the scaling of the energy of excited states $E_i$ with the polymer length $t$. In Fig.~\ref{fig:betazeta} we show the level $i$ energy and end-point fluctuations $\sigma^2_{E,i}(t) = \langle E_i^2\rangle - \langle E_i \rangle^2 \sim t^{2\beta}$ and $\sigma^2_{|x|,i}(t) = \langle |x_i|^2\rangle - \langle |x_i| \rangle^2 \sim t^{2/z}$ scaling with polymer length $t$ for different values of $i$. For low energy levels, the scaling corresponds to the values of $\beta$ and $z$ of KPZ, as it does for the ground state at $i=0$. However, for large enough energy levels, the scaling crosses over to new values $\beta\simeq0.17$ and $z\simeq0.8$. 
Note also that the value of $\delta$ measured for energy levels with large gap is quite close to that which corresponds from the scaling exponents measured in Fig.~\ref{fig:Sk}, $\delta = 2\beta z +1 \approx 1.136$, hence scaling seems to be consistent.

%%%%%%%%%%%%%%%%%%%%%%%%%%%%%%%%%%%%%%%%%%%%% FIG3

\paragraph{Discussion.-} The equivalence between the stochastic-field equations that describe the time evolution of infinitesimal perturbations in space-time chaos, Eq.~(\ref{mult_noise_eq}), and the DPRM problem, Eq.~(\ref{partition}), can be exploited to construct a mapping between both problems. Indeed, the 1st CLV amplitude $|g_1(x,t)|$ is formally equivalent to the partition function $Z(x,t)$ of the DPRM. Both are known to become strongly localized in space. For the polymer the partition function localizes at the ground state path, $Z(x,t) \sim \exp[-E_0t - g(x/t^{1/z})]$, where $E_0$ is the minimum energy and $g$ is a scaling function describing the sample-to-sample transverse fluctuations of the end-point~\cite{Kardar.Zhang_1987}. Conversely, for the main CLV, $-E_0$ is nothing but the 1st LE and $g$ describes its fluctuations. Note that fluctuations of the main LE were theoretically proven to be described by KPZ universality ($z=3/2$)~\cite{plp13}. Other solutions of the multiplicative-noise equation are unattainable because their basin of attraction has zero measure: For any given noise realization the system gets trapped in the corresponding ground state. The $n$-th CLV can only be obtained by actively excluding the subspace spanned by the first $n-1$ CLVs, which is equivalent to excluding the first $n-1$ energy levels in the DPRM problem in order to get the polymer path corresponding to the $n$-th energy level. 

Our numerical simulations show that the excited energy states of the $1+1$ DPRM at zero temperature exhibit a non trivial sample-to-sample scaling for long enough polymer lengths $t$. The spatial distribution of energy fluctuations shows a correlation that crosses over from a KPZ $\delta = 2$ exponent at short distances to $\delta = 1.3$ at large scales. The partition function, if restricted to paths with energies $E > E_i$ for $i \gg 0$, should be $Z_i(x,t) \sim \exp[-E_it - \tilde{g}(x/t^{1/z'})]$, where $E_i$ is the energy of the $i$-th excited state, $\tilde{g}$ a scaling function, and $z' \approx 0.8$. Note that the new scaling regime appears for energy correlations between high energy levels, which implies long enough distances $x \sim t$ (taking into account the semielliptic profile in Fig.~\ref{fig:profile}). This simple heuristic argument gives $z' \approx 1$.  Obviously, more precise analytical arguments would be required to get better estimates. 
Correspondingly, the $n$-th CLV would behave asymptotically as  $|g_n(x,t)| \sim \exp[\lambda_n t + \tilde{g}(x/t^{1/z'})]$, where $\lambda_n$ is the $n$-th LE, which is consistent with the scaling behavior observed in numerical simulations of dissipative chaos in extended systems~\cite{szendro07,pazo08}. We stress here that in order to obtain from theory the crossover to $\delta \approx 1.3$ that we observed in our simulations one would require to calculate two-point correlations of the type $\langle \ln Z(x,t) \ln Z(x',t) \rangle$ not only for the ground state, but for high energy levels. As far as we know this has not been done so far.

In summary, we have found a non trivial exponent $\delta \approx 1.3$ that describes the long wavelength behavior of spatial correlations between sample-to-sample fluctuations of excited energy levels in the DPRM. This exponent can be identified with that observed in the problem of the scaling of bulk CLVs in spatio-temporal chaos~\cite{szendro07,pazo08}. In the field of space-time chaos this exponent has been shown to be essential to explain the universal scaling of the fluctuations of bulk LEs~\cite{plp13,plp16}. Our results can open the door to further theoretical insights on the origin of this exponent by using the analytical tools developed in recent years to treat the DPRM based on the formulation of the problem in terms of a gas of attractive bosons, the Bethe ansatz, the replica trick, or random matrix theory~\cite{Kardar1987,dotsenko_2000,Calabrese2011,Borodin2016,Dotsenko2017,Dotsenko_2022}.
 
Finally, we stress that ensembles of optimal paths like those that appear here are very common in statistical mechanics and condensed matter physics. Indeed, similar path structures appear in river basins, traffic networks, interface growth, fracture cracks, as well in many other problems in random systems. We do believe that a complete understanding of the unexpected scaling behavior of the exited energy states that we report here can provide important insights into other seemingly unrelated random systems.

\acknowledgments

This work was supported by Grant No.~PID2021-125543NB-I00 funded by MCIN/AEI/10.13039/501100011033/ and by ERDF “A way of making Europe” by the European Union. ER acknowledges support from Margarita Salas postdoctoral program from Universidad Carlos III Madrid. 

\bibliography{lv_dprm}

\end{document}